\documentclass{article}

\usepackage[preprint]{neurips_2026}

\usepackage[utf8]{inputenc}
\usepackage[T1]{fontenc}
\usepackage{url}
\usepackage{booktabs}
\usepackage{xspace}
\usepackage{amsfonts}
\usepackage{amsthm}
\usepackage{amsmath}
\usepackage{amssymb}
\usepackage{nicefrac}
\usepackage{microtype}
\usepackage{xcolor}
\usepackage{stmaryrd}
\usepackage{algorithm}
\usepackage{algorithmic}
\usepackage{graphicx}
\usepackage{subcaption}
\usepackage{enumitem}
\usepackage{float}
\usepackage{pgf}
\usepackage{import}

\usepackage{hyperref}
\usepackage[capitalize]{cleveref}

\newfloat{eqfloat}{htbp}{eqf}

\floatname{eqfloat}{Equation}
\crefname{eqfloat}{Equation}{Equations}
\Crefname{eqfloat}{Equation}{Equations}

\newcommand{\SIMBA}{\textsc{Simba}\xspace}
\newcommand{\VFB}{\textsc{VFB}\xspace}
\newcommand{\cvc}{\textsc{cvc5}\xspace}
\newcommand{\SIMBAabl}{\textsc{Simba-RTid}\xspace}
\newcommand{\bv}[1]{\mathbf{#1}}
\newcommand{\N}{\mathbb{N}}

\newcommand{\B}{\mathbb{B}}
\newcommand{\cost}{\mathsf{size}}

\newcommand{\band}{\mathbin{\land}}
\newcommand{\bor}{\mathbin{\lor}}
\newcommand{\bxor}{\mathbin{\oplus}}
\newcommand{\bnot}{\mathord{\sim}}
\newcommand{\bneg}{\mathord{-}}

\title{GPU-Accelerated Synthesis\\ of Mixed-Boolean Arithmetic: \\Beyond Caching}

\author{
  Gabriel Bathie, Baptiste Mouillon, and Nathanaël Fijalkow\\
  CNRS, LaBRI, University of Bordeaux, France
}

\theoremstyle{definition}
\newtheorem{definition}{Definition}[section]

\begin{document}

\maketitle

\begin{abstract}
    Synthesizing Mixed-Boolean Arithmetic (MBA) expressions from input-output examples is central to program deobfuscation and also useful for compiler optimization, reverse engineering, and cryptanalysis.
  Existing MBA synthesizers are typically CPU-based and scale poorly on large specifications or complex targets.
  Recent GPU-accelerated synthesis methods achieve large speedups in qualitative settings, but they depend on caching observationally equivalent candidates; this strategy breaks down for MBA because candidate outputs are quantitative bitvectors and the behavioral space is enormous.
  We present \SIMBA{} (\textbf{S}ynthesis of \textbf{M}ixed-\textbf{B}oolean \textbf{A}rithmetic), a GPU-accelerated MBA synthesizer built around cache-free bottom-up enumeration.
  \SIMBA{} avoids language caches entirely and uses a GPU-oriented enumeration design that keeps work local and highly parallel.
  In experiments, \SIMBA{} is substantially faster than prior MBA synthesis tools, handles larger specifications, and reaches expression sizes that existing methods fail to solve.
  These results establish cache-free GPU synthesis as a practical and scalable approach for quantitative domains, and identify it as a strong alternative to cache-centric designs.

\end{abstract}

\section{Introduction}
\label{sec:intro}
Program synthesis, the automated generation of programs from specifications, is a fundamental challenge in computer science~\citep{gulwani2017program, david2017program}.
Among the many formulations of program synthesis, \emph{synthesis from input-output examples} (also known as inductive programming or programming from example) is both practically important and theoretically challenging: given a set of example pairs $\{(\bv{x}_i, y_i)\}_{i=1}^{n}$, find a program $e$ in a target language such that $e(\bv{x}_i) = y_i$ for all~$i$.

In this paper, we study \emph{Mixed-Boolean Arithmetic} (MBA) synthesis from examples.
MBA expressions combine bitwise operations (AND $\band$, OR $\bor$, XOR $\bxor$, NOT $\bnot$) with arithmetic operations (addition $+$, subtraction $-$, multiplication $\times$) over fixed-width bitvectors.
This problem arises in several important domains.
The primary application is \emph{code deobfuscation}: MBA obfuscation~\citep{zhou2007information, eyrolles2017obfuscation} transforms simple expressions (e.g., $x + y$) into semantically equivalent but syntactically complex ones (e.g., $(x \bxor y) + 2 \cdot (x \band y)$), and recovering the original is naturally cast as MBA synthesis~\citep{blazytko2017syntia, david2020qsynth, lee2023simplifying, liu2021mba, reichenwallner2023gamba, attias2025augmenting}.
MBA synthesis also serves compiler superoptimizers~\citep{bansal2006superoptimizer, sasnauskas2017souper}, which search for simpler equivalent instruction sequences, and supports reverse engineering and cryptanalysis by extracting human-readable descriptions of complex binary computations.

\medskip

Current MBA synthesis methods do not scale well. Stochastic systems such as Syntia~\citep{blazytko2017syntia} and its successors Xyntia~\citep{MenguyBBL21a} and XSmir~\citep{attias2025augmenting} can be effective in practice but do not provide completeness guarantees and struggle on hard instances.
Algebraic simplifiers such as MBA-Blast~\citep{liu2021mba}, SiMBA~\citep{reichenwallner2022efficient}, and GAMBA~\citep{reichenwallner2023gamba} are strong on important subclasses but do not cover the full synthesis setting.
SAT/SMT-based methods are principled but expensive at larger bit-widths, and neural approaches such as NeuReduce~\citep{feng2020neureduce} trade soundness for speed.

A natural idea is to bring MBA synthesis to GPUs.
Bottom-up enumeration is massively parallel, and prior work on regular-expression inference~\citep{valizadeh2023paresy} and LTL learning~\citep{valizadeh2024ltl} shows that GPU acceleration can deliver major speedups.
But these methods rely on a language cache indexed by observational equivalence.
That is effective in qualitative domains, where behavior on examples is Boolean.
In MBA synthesis, behavior is quantitative: each candidate returns a $w$-bit value per example.
With $n$ examples, the behavioral space has size $2^{wn}$ (e.g., $2^{320}$ for $n=10$, $w=32$), so cache deduplication quickly becomes ineffective due to memory pressure and collisions.

Our core claim is that GPU acceleration for MBA requires moving beyond cache-centric designs.
We introduce \SIMBA{}, a cache-free GPU MBA synthesizer based on bottom-up enumeration by expression size.
\SIMBA{} is built around the following ideas:
\begin{enumerate}
  \item \textbf{Cache-free on-the-fly evaluation.} Each candidate expression is decoded from a thread ID, evaluated against the specification, and immediately discarded. No intermediate results are stored, so the memory bottleneck of language caches disappears entirely.
  \item \textbf{Bijective enumeration.} We define a bijection between integers and MBA expressions of a given size. Each GPU thread is assigned a unique integer, independently decodes it into an expression, and evaluates it with no inter-thread communication.
  \item \textbf{Locally consistent encoding.} The bijection is designed so that consecutive integers decode to structurally similar expressions. Threads within the same GPU warp therefore follow nearly identical execution paths, minimizing branch divergence and keeping hardware utilization high.
\end{enumerate}

Empirically, \SIMBA{} synthesizes expressions substantially faster than prior MBA tools, handles larger specifications, and reaches larger target sizes.
The main takeaway is that quantitative synthesis can benefit strongly from GPUs, provided the algorithm is designed to avoid cache bottlenecks from the start.

\section{Problem Definition}
\label{sec:problem}
\subsection{Bitvectors and MBA Expressions}

We work with $w$-bit bitvectors, i.e., elements of $\B^w = \{0, 1\}^w$, which we also identify with integers in $\{0, \ldots, 2^w - 1\}$ (unsigned) or $\{-2^{w-1}, \ldots, 2^{w-1} - 1\}$ (signed) via the standard binary encoding.
In practice, we use $w = 32$.

\begin{definition}[MBA expressions]
  \label{def:mba}
  Let $\mathcal{V} = \{x_1, \ldots, x_k\}$ be a set of $k$ variables ranging over $\B^w$.
  The set $\mathsf{MBA}(\mathcal{V})$ of \emph{Mixed-Boolean Arithmetic expressions} over $\mathcal{V}$ is defined by the grammar:
  \begin{align*}
    e \;::=\;\; & x_i \in \mathcal{V} \mid \bneg e \mid \bnot e \mid e_1 \band e_2 \mid e_1 \bor e_2 \mid e_1 \bxor e_2 \mid e_1 + e_2 \mid e_1 - e_2 \mid e_1 \times e_2.
  \end{align*}

\end{definition}

All operations are over $\B^w$, i.e., arithmetic is modular ($\bmod\; 2^w$) and bitwise operations act independently on each bit position.
We write $\llbracket e \rrbracket : (\B^w)^k \to \B^w$ for the \emph{semantics} of expression $e$, defined in the standard way.

\begin{definition}[Size]
  \label{def:size}
  The \emph{size} of an MBA expression $e$, written $\cost(e)$, counts the number of operators and variables. In other words, it is the size of the tree representing the expression $e$.

\end{definition}

\subsection{The MBA Synthesis Problem}

\begin{definition}[Specification]
  A \emph{specification} is a set $S = \{(\bv{x}_1, y_1), \ldots, (\bv{x}_n, y_n)\}$ of input-output pairs where each $\bv{x}_i = (x_{i,1}, \ldots, x_{i,k}) \in (\B^w)^k$ is a $k$-tuple of $w$-bit bitvector inputs and $y_i \in \B^w$ is the expected output.
  We require that the inputs are pairwise distinct: $\bv{x}_i \neq \bv{x}_j$ for $i \neq j$.
\end{definition}

\begin{definition}[MBA Synthesis Problem]
  \label{def:synthesis}
  Given a specification $S = \{(\bv{x}_i, y_i)\}_{i=1}^{n}$ and a size budget $C \in \N$:
  \begin{itemize}[noitemsep,topsep=0pt]
    \item \textbf{Soundness:} Find an expression $e \in \mathsf{MBA}(\mathcal{V})$ such that $\llbracket e \rrbracket(\bv{x}_i) = y_i$ for all $i \in \{1, \ldots, n\}$.
    \item \textbf{Minimality:} Among all sound expressions, find one of minimum size: $\cost(e) \leq \cost(e')$ for all sound $e'$.
    \item \textbf{Bounded synthesis:} Find a sound expression with $\cost(e) \leq C$, or report that none exists.
  \end{itemize}
\end{definition}

\paragraph{Observational equivalence.}
Two expressions $e_1, e_2$ are \emph{observationally equivalent} with respect to specification $S$, written $e_1 \equiv_S e_2$, if $\llbracket e_1 \rrbracket(\bv{x}_i) = \llbracket e_2 \rrbracket(\bv{x}_i)$ for all $i$.
The observational behavior of $e$ given $S$ is the vector $\mathsf{obs}_S(e) = (\llbracket e \rrbracket(\bv{x}_1), \ldots, \llbracket e \rrbracket(\bv{x}_n)) \in (\B^w)^n$.

\paragraph{Qualitative vs.\ quantitative.}
In prior GPU-based synthesis for LTL~\citep{valizadeh2024ltl}, observational behavior is a Boolean vector in $\B^n$, since each formula either accepts or rejects each trace.
With $n = 64$ traces, there are only $2^{64}$ possible behaviors, fitting in a single 64-bit machine word.
In MBA synthesis, $\mathsf{obs}_S(e) \in (\B^w)^n$, giving $2^{wn}$ possible behaviors.
With $n = 10$ examples and $w = 32$, this is $2^{320}$, which is utterly infeasible to cache.
This is the fundamental reason why the language-cache approach of~\citet{valizadeh2024ltl, valizadeh2023paresy} cannot be applied to MBA synthesis, and new algorithmic ideas are required.

\section{Algorithm}
\label{sec:algorithm}
We now describe the \SIMBA{} algorithm. The high-level structure is a bottom-up enumeration of MBA expressions by increasing size. The key novelty is that, unlike prior GPU-based synthesizers, \SIMBA{} does not use a language cache to store and deduplicate intermediate results, but instead relies on carefully crafted encoding of MBA expressions.

\subsection{Cache-Free Enumeration}
\label{sec:cache-free}

The key idea of \SIMBA{} is to replace the language cache entirely with a \emph{bijection} between integers and MBA expressions.
Each GPU thread is assigned a unique integer identifier, decodes it into an expression on-the-fly, evaluates that expression against the specification, and then discards it.
No intermediate results are stored between threads, and no inter-thread communication is needed beyond a final atomic write when a solution is found.

\paragraph{Expression representation.}
Expressions are represented in \emph{Reverse Polish Notation} (RPN).
A formula of size $s$ is a sequence of $s$ tokens from the alphabet $\{x_0, \ldots, x_{k-1}\} \cup \{\bnot, \bneg, \band, \bor, \bxor, +, -, \times\}$, where binary operators pop two operands and push one result, and unary operators pop one and push one.
Variables are the leaves (size 1), and operators increase size by exactly one (unary) or combine two sub-expressions (binary).
The supported operator set consists of two unary operators ($\bnot$ and $\bneg$) and six binary operators ($\band$, $\bor$, $\bxor$, $+$, $-$, $\times$), for a total of 8 operator slots.

\begin{eqfloat}
  \noindent
  \begin{minipage}[htbp]{.5\linewidth}
    \begin{align*}
      \mathsf{T}[1][8] &= k \quad (\text{one expression per variable}),\\
      \mathsf{T}[s][\mathit{op}_{\text{unary}}] &= \mathsf{T}[s{-}1][8], \\
      \mathsf{T}[s][\mathit{op}_{\text{comm}}] &= \sum_{j=1}^{\lfloor (s-1)/2 \rfloor} \mathsf{T}[j][8]\cdot \mathsf{T}[s{-}1{-}j][8],
    \end{align*}
  \end{minipage}
  \begin{minipage}[htbp]{.5\linewidth}
    \begin{align*}
      \mathsf{T}[s][\mathit{op}_{\text{sub}}] &= \sum_{j=1}^{s-2} \mathsf{T}[j][8]\cdot \mathsf{T}[s{-}1{-}j][8], \\
      \mathsf{T}[s][8] &= 2\,\mathsf{T}[s][\mathit{op}_{\text{unary}}] \\
        &\quad + 5\,\mathsf{T}[s][\mathit{op}_{\text{comm}}] + \mathsf{T}[s][\mathit{op}_{\text{sub}}],
    \end{align*}
  \end{minipage}
  \caption{Recurrence for the table $\mathsf{T}[s][\mathit{op}]$.}\label{eqtable}
\end{eqfloat}

\paragraph{Counting table.}
The bijection is enabled by a precomputed table $\mathsf{T}[s][\mathit{op}]$ counting the number of valid RPN expressions of size $s$ whose top-level operator is $\mathit{op}$ (with $\mathsf{T}[s][8]$ denoting the total across all operators).
The table is computed on the CPU by the recurrence described in \cref{eqtable},
where commutative binary operators ($\band$, $\bor$, $\bxor$, $+$, $\times$) count only ordered pairs with $|L| \leq |R|$ to avoid duplicates, while the non-commutative subtraction ($-$) counts all ordered pairs.
Before enumeration begins, the table is transferred to the so-called \emph{constant} memory of the GPU, a small segment of the device's memory that is read-only to threads but can be accessed with minimal latency.

\paragraph{Bijection and decoding.}
For each size $s$, the integers $\{0, \ldots, \mathsf{T}[s][8]-1\}$ are in bijection with all valid RPN expressions of size $s$ via the decoding function $\phi_s$:
\begin{itemize}[noitemsep,topsep=0pt]
  \item \textbf{Base case} ($s = 1$): $\phi_1(n) = x_n$.
  \item \textbf{Find top operator}: scan the cumulative sums $\mathsf{T}[s][0], \mathsf{T}[s][0]+\mathsf{T}[s][1], \ldots$ until the prefix exceeds $n$; call this operator $\mathit{op}$, and let $n' = n - \text{(sum before } \mathit{op}\text{)}$.
  \item \textbf{Unary} ($\mathit{op} \in \{\bnot, \bneg\}$): $\phi_s(n) = \mathit{op}(\phi_{s-1}(n'))$.
  \item \textbf{Binary}: find the left-subtree size $j$ such that $n'$ falls in the block for split $(j, s{-}1{-}j)$, i.e., scan cumulative sums of $\mathsf{T}[j][8]\cdot\mathsf{T}[s{-}1{-}j][8]$ over valid $j$; let $n'' = n' - \text{offset}$. Then $\phi_s(n) = \phi_j\!\left(\lfloor n'' / \mathsf{T}[s{-}1{-}j][8] \rfloor\right) \;\mathit{op}\; \phi_{s-1-j}\!\left(n'' \bmod \mathsf{T}[s{-}1{-}j][8]\right)$.
\end{itemize}
This decoding is computed entirely in thread-local registers; no global memory reads beyond the constant-memory table.

\paragraph{Enumeration.}
The enumeration is shown in Algorithm~\ref{alg:simba}.
For each size $s$ and each operator slot $\mathit{op}$, the host launches a GPU kernel with exactly $\mathsf{T}[s][\mathit{op}]$ threads.
Thread $\mathsf{tid}$ decodes $\phi_s(\mathsf{tid} + \mathsf{offset}_{\mathit{op}})$ where $\mathsf{offset}_{\mathit{op}} = \sum_{\mathit{op}' < \mathit{op}} \mathsf{T}[s][\mathit{op}']$ is the cumulative count of preceding operator slots, evaluates the resulting expression against all $n$ input-output pairs using an RPN stack interpreter, and checks if outputs match.
If so, it atomically claims the solution slot and writes the formula.
The host reads the solution buffer after each operator batch.

\begin{algorithm}[htbp]
  \caption{\SIMBA{}: bijective cache-free enumeration of MBA expressions}
  \label{alg:simba}
  \begin{algorithmic}[1]
    \REQUIRE Specification $S = \{(\bv{x}_i, y_i)\}_{i=1}^n$, $k$ variables, size bound $C$
    \ENSURE Expression $e$ with $\llbracket e \rrbracket(\bv{x}_i) = y_i$ for all $i$, or \textsf{Not found}
    \STATE Compute counting table $\mathsf{T}$ on CPU; copy to GPU constant memory
    \FOR{$s = 1, 2, \ldots, C$}
    \STATE $\mathsf{offset} \leftarrow 0$
    \FOR{$\mathit{op} \in \{\bnot,\, \band,\, \bor,\, \bxor,\, \bneg,\, +,\, -,\, \times\}$}
    \FORALL{$\mathsf{tid} = 0, \ldots, \mathsf{T}[s][\mathit{op}]{-}1$ \textbf{in parallel on GPU}}
    \STATE $e \leftarrow \phi_s(\mathsf{tid} + \mathsf{offset})$ \quad \COMMENT{decode integer to RPN expression}
    \IF{$\llbracket e \rrbracket(\bv{x}_i) = y_i$ for all $i$}
    \STATE Atomically write $e$ to solution buffer
    \ENDIF
    \ENDFOR
    \STATE Check solution buffer on host; \textbf{return} $e$ if found
    \STATE $\mathsf{offset} \leftarrow \mathsf{offset} + \mathsf{T}[s][\mathit{op}]$
    \ENDFOR
    \ENDFOR
    \RETURN \textsf{Not found}
  \end{algorithmic}
\end{algorithm}

\subsection{GPU-friendly Parallel Enumeration}

\paragraph{GPU Architecture.}
Modern GPUs can execute operations in parallel on thousands of threads.
However, they cannot execute \emph{arbitrary} operations in parallel like CPUs; instead, they follow a paradigm called \emph{Single Instruction, Multiple Threads} (SIMT)~\citep{nickolls2008scalable}.
SIMT shares similarities with traditional SIMD (Single Instruction, Multiple Data) vector organizations by broadcasting a single instruction to multiple execution units (\emph{threads}) simultaneously to maximize hardware throughput.
Additionally, the SIMT architecture can handle some degree of divergence, for example when threads of the same hardware group (called a \emph{warp}) take different branches of an \texttt{if-else} statement.
In this situation, threads that enter the \texttt{if} branch are executed first, while threads assigned to the \texttt{else} branch are temporarily masked out and remain idle.
Once the first branch completes, the hardware inverts the execution mask to run the \texttt{else} branch. This process serializes the execution of the divergent paths, temporarily reducing the effective instruction throughput, until all threads in the warp complete the conditional block and execution reconverges.

Therefore, to maximize parallelism, one must ensure that threads of the same wrap say in sync as much as possible and do not take divergent branches.
In our case, this means that threads of the same warp must evaluate \emph{structurally similar} MBA expressions.

\paragraph{Locally Consistent Encoding of MBA Expressions.}
Each execution thread is given a unique integer identifier from $[0, n_{threads})$, with each warp being assigned identifiers from a contiguous slice of that range.
In our algorithm, each thread is assigned a single MBA expression based on its identifier, and checks whether it satisfies the specification.
We use a bijective mapping from integers to MBA expressions of a given size with the following properties:
\begin{itemize}
  \item all threads are assigned different expressions (injectivity),
  \item all expressions are assigned to some thread (surjectivity),
  \item threads within the same warp process very similar formulas: most of the time, they differ only at the input variables.
\end{itemize}
This encoding allows us to leverage the full power of the GPU by ensuring no redundant work is done and minimizing idle threads due to branch divergence.
Furthermore, the ensure that decoding an integer to the corresponding MBA expression can be computed efficiently, in parallel.

\paragraph{Concrete mapping.}
The bijection $\phi_s$ is locally consistent by construction.
Consider a kernel batch for a binary operator at size $s$ with a fixed left-subtree size $j$.
Within this batch, thread $\mathsf{tid}$ decodes to left sub-expression $\phi_j\!\left(\lfloor \mathsf{tid} / \mathsf{T}[s{-}1{-}j][8] \rfloor\right)$ and right sub-expression $\phi_{s-1-j}\!\left(\mathsf{tid} \bmod \mathsf{T}[s{-}1{-}j][8]\right)$.
Hence, consecutive thread IDs share the same left sub-expression and differ only in the right sub-expression.
Within a warp of 32 threads, all threads therefore:
\begin{enumerate}
  \item evaluate the \emph{same} left sub-expression, executing the same decode path and producing the same intermediate values, ensuring no branch divergence;
  \item evaluate consecutive right sub-expressions, whose decoding traces differ only in the final variable index which again, ensures minimal divergence.
\end{enumerate}
For unary operators, consecutive threads decode expressions that differ only in the last token of the child sub-expression, which similarly minimizes divergence within a warp.

\section{Experiments}
\label{sec:experiments}
We evaluate \SIMBA{} along three axes: (1)~comparison with existing MBA synthesis tools on a new, challenging benchmark suite; (2)~a quantitative analysis of why cache-based approaches fail to scale, and how \SIMBA{}'s cache-free design overcomes this bottleneck; and (3)~an ablation study isolating the contribution of the locally consistent encoding.

\subsection{Experimental Setup}

\paragraph{Benchmarks.}

The existing benchmark suite proposed in~\cite{MenguyBBL21a} turned out too easy for \SIMBA, as it solved 90\% of the instances in less than a second. We therefore created a new benchmark suite of 2340 randomly generated MBA expressions following the exact same generation procedure with larger parameters. Three parameters determine the difficulty of an MBA synthesis problem: the expression size ($s$), the number of variables ($k$), and the number of examples in the specification ($n$).

To build the dataset, we generate 10 random expressions for every combination of $s = 5, 6, \dots, 30$ and $k = 2, \dots, 10$.
The generated expressions might not be minimal: there may be smalle equivalent formulas, and they may depend on fewer than $k$ variables. For example, if $k=4$, the expression $e = x_2 + (x_0 \band x_0)$ has size 5 and simplifies to $e' = x_2 + x_0$, of size 3, and actually uses only two variables.
To accurately measure an instance's difficulty, we redefine its size based on the smallest known formula (comparing our generated expression against the solver's eventual solutions). We then record its true variable count by taking the highest variable index present in that minimal formula, plus one. The above example would therefore be reported with $s = 3$ and $k = 3$ (the highest variable index is 2).

\begin{figure}[htbp]
  \begin{centering}
    \import{figures/}{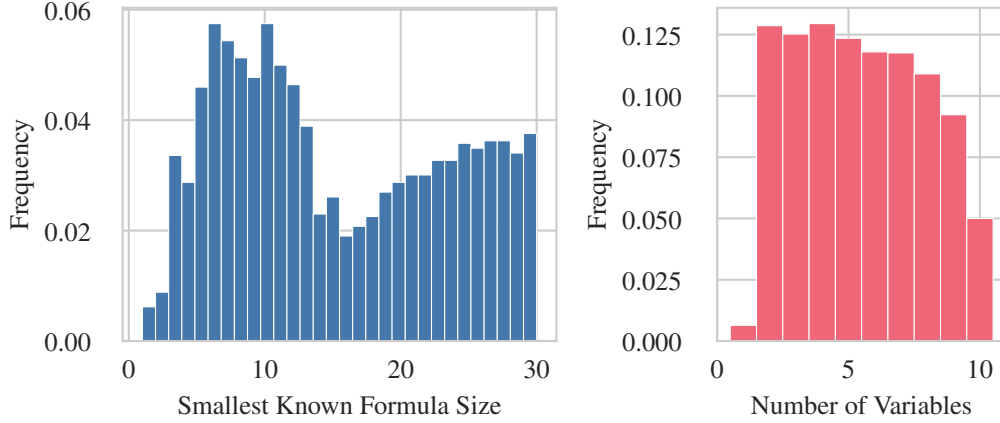}
    \caption{Distributions of formula size and number of variables across the benchmark suite.}\label{fig:benchmark_distributions}
  \end{centering}
\end{figure}

The distributions of size and number of variables across our benchmark suite are shown in \cref{fig:benchmark_distributions}, covering a wide range of problem sizes and complexities.

Finally, we construct specifications by evaluating each expression on $n=16$ examples, using uniformly random 32-bit values for each input. This value of $n$ ensures that it is statistically unlikely that a non-equivalent formula produces identical outputs, guaranteeing that any solution to the synthesis problem is strictly equivalent to the original formula.

\paragraph{Baselines.}
We compare \SIMBA against the following state-of-the-art exact MBA synthesis solvers:
\begin{itemize}
  \item \textbf{\VFB}: an implementation of the GPU algorithm of \citet{valizadeh2024ltl} for MBA instead of LTL formulas.
  \item \textbf{cvc5}~\citep{cvc5}: a CPU SMT-based SyGuS solver (bitvector theory).

\end{itemize}

\paragraph{Hardware.}

All CPU and H100 GPU experiments were conducted on nodes equipped with two Intel Xeon Platinum 8468 CPUs (48 cores @2.10 GHz each), 512 GB RAM, and eight NVIDIA H100 SXM5 GPUs with 80 GB of memory each.
Experiments on A100 GPUs were conducted on nodes equipped with two AMD Milan EPYC 7543 CPUs (32 cores @2.80 GHz each), 512 GB RAM and eight NVIDIA A100 SXM4 GPUs with 80 GB of memory each.
We report wall-clock times for solving each instance, excluding the overhead for parsing the input.
Results on A100 GPUs and a stability analysis over 5 runs are reported in \cref{app:hardware}; they show consistent behaviour across both hardware platforms and very little run-to-run variance.

\subsection{Comparison with Existing Tools}

\begin{figure}[htbp]
  \begin{centering}
    \import{figures/}{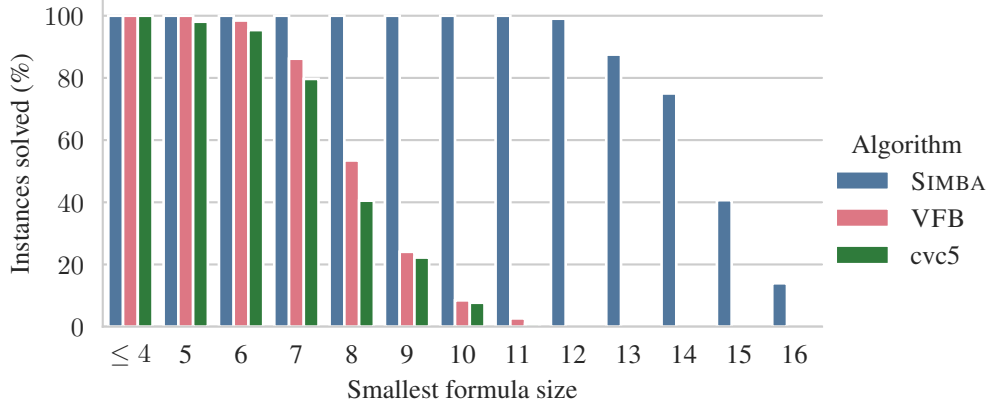}
    \caption{Percentage of instances solved by each algorithm across varying formula sizes.}\label{fig:sizes}
  \end{centering}
\end{figure}

We run all three solvers (\SIMBA, \VFB and \cvc) on our dataset, with a time limit of 60 seconds.
We report the results in \Cref{fig:sizes}. \SIMBA solves nearly all instances up to size 11 and is the only algorithm capable of solving instances of sizes 12 through 16. In contrast, the performance of \VFB and \cvc degrades rapidly after size 6, and cannot solve instances of size greater than 11. No program solved instances of size more than 16. Similar conclusions can be drawn observing the influence on the number of variables, see \cref{app:comparison_existing_tools}.

In addition to the coverage gap, a head-to-head comparison of solving times reveals two further patterns.
Against \VFB: among the 615 instances solved by both, \SIMBA is faster on 90\% of them, with a median speedup of $1.6\times$.
The remaining 634 instances solved exclusively by \SIMBA are entirely out of reach for \VFB.
Against \cvc: on the 581 instances both solvers handle, \SIMBA finishes in under one second on 96\% of them (median: $0.67$\,s), indicating that they are rather simple instances.
The remaining 668 instances solved only by \SIMBA are beyond \cvc's reach within the time limit.

\subsection{Scalability}

The scalability of \SIMBA stems from its cache-free design. \VFB maintains a language cache of all intermediate sub-expressions encountered during enumeration: each new expression of size $s$ is built from cached sub-expressions of smaller sizes. This enables deduplication of semantically equivalent sub-expressions, but at the cost of memory that grows super-exponentially with formula size.

\paragraph{Memory scaling of cache-based approaches.}
\Cref{tab:cache-5} shows, for $k = 5$ variables, how the number of cached formulas and the resulting memory footprint grow with the enumeration size. \VFB stores the observational behaviour of all expressions in the cache; it exhausts GPU memory already at size~11. Below that threshold, the cache reaches 2.1~GB at size~10. Memory becomes the limiting factor long before the solver can tackle the instances that \SIMBA handles routinely.

\paragraph{Diminishing returns of deduplication.}
\Cref{tab:cache-5} also shows the fraction of all possible formulas actually stored in the cache, as well as the cumulative median time \VFB spends to reach each size. At size~2 the cache is complete (100\%), but the proportion falls rapidly as the search space grows exponentially. By size~11 the cache would need to store over 4~billion formulas, crashing the GPU long before that point. At size~10 only 7.4\% of the 439~million candidate formulas are cached, yet those 32~million entries already occupy 2.1~GB. The exponential growth of the full search space outpaces the pruning provided by deduplication, making cache-based approaches fundamentally memory-bound.

\begin{table}[htbp]
\centering
\caption{Cache size and efficiency of \VFB for $k = 5$ variables. \VFB runs out of GPU memory (OOM) once the number of stored formulas is too large, making it unable to search larger formula sizes.}
\label{tab:cache-5}
\begin{tabular}{rrrrrr}
\toprule
Size & \#MBA & \#VFB cache & VFB mem & \% cached & VFB cum.\ time (s) \\
\midrule
2 & 15 & 15 & $<$1\,MB & 100.0\% & 1.2 \\
3 & 185 & 99 & $<$1\,MB & 53.5\% & 2.4 \\
4 & 875 & 166 & $<$1\,MB & 19.0\% & 3.4 \\
5 & 8,805 & 1,749 & $<$1\,MB & 19.9\% & 4.3 \\
6 & 60,715 & 6,874 & $<$1\,MB & 11.3\% & 5.5 \\
7 & 663,785 & 115,080 & 7.4\,MB & 17.3\% & 6.5 \\
8 & 5,062,975 & 504,522 & 32.3\,MB & 10.0\% & 7.4 \\
9 & 50,895,805 & 5,547,921 & 355.1\,MB & 10.9\% & 8.5 \\
10 & 438,822,815 & 32,523,385 & 2.1\,GB & 7.4\% & 9.3 \\
11 & 4,472,457,185 & OOM & OOM & OOM & OOM \\
\bottomrule
\end{tabular}
\end{table}

\paragraph{Cache-free scalability.}
\SIMBA requires no cache at all: each thread independently decodes and evaluates its assigned expression on-the-fly, then discards it. The memory footprint is therefore constant with respect to formula size, bounded only by the GPU's register file and constant memory (used for a small pre-computed table of expression counts). This lets \SIMBA enumerate expressions of sizes that would be inaccessible to cache-based approaches under any practical memory budget. The results in \cref{fig:sizes} confirm this: \SIMBA is the only solver capable of solving instances of sizes 12 through 16.

Results for other variable counts ($k \in \{3, 4, 6, 7, 8\}$) follow the same pattern and are given in \cref{app:cache-tables}.

\subsection{Ablation Study}

\paragraph{Effect of the Locally Consistent Encoding.}
Recall that \SIMBA assigns each GPU thread a unique integer identifier, which it decodes on-the-fly into an MBA expression.
A key design principle of this encoding is \emph{local consistency}: consecutive thread identifiers decode to structurally similar expressions.
Since threads within the same GPU warp share a contiguous range of identifiers, they evaluate expressions that differ only in their input variables, following nearly identical execution paths.
This minimizes branch divergence within the warp and keeps the GPU hardware fully utilized.

\textbf{How much does this design choice actually matter?}
To isolate its effect, we introduce \SIMBAabl, a variant of \SIMBA in which thread identifiers are pseudo-randomly permuted before decoding.
This shuffling preserves the coverage of the search space: every expression is still evaluated exactly once, but destroys the locality structure since threads within the same warp now process structurally diverse expressions, inducing significant branch divergence.

\begin{figure}[htbp]
  \centerline{\resizebox{.8\columnwidth}{!}{\input{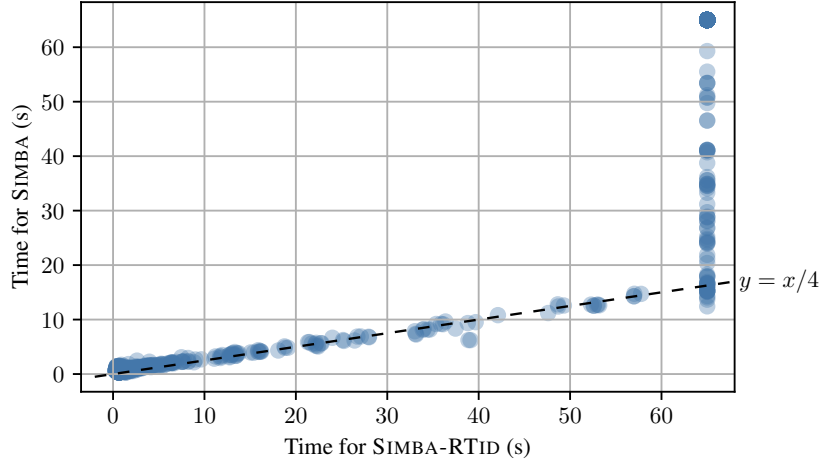}}}
  \caption{Comparison of the solving time of \SIMBAabl and \SIMBA on each instance. Each circle represents an instance; its $x$-coordinate is the solving time for \SIMBAabl and its $y$-coordinate the time for \SIMBA.
  The lines $x=65$ and $y = 65$ correspond to timeouts of \SIMBAabl and \SIMBA, respectively. Most instances solved by both programs lie on the $y = x/4$ line, which means that \SIMBA is 4 times faster than \SIMBAabl.}\label{fig:ablation}
\end{figure}

To permute the threads, we use simple modular arithmetic: the thread with original ID $i$ gets assigned the ID $\pi(i)$, where
\[\pi(i)= (i \cdot 2246822507) \mod n_{threads}.\]
The constant $C = 2246822507$ is taken from the MurmurHash3 hash function~\citep{murmurhash}. Since the number of threads is always even, $C$ is coprime with $n_{threads}$, and $\pi$ is indeed a permutation over $\{0, \ldots, n_{threads}-1\}$.
The results of this experiment are reported in \cref{fig:ablation}.
We observe that \SIMBA solves all instances that \SIMBAabl solves, and is consistently four times faster in doing so.

\section{Related Work}
\label{sec:related}
\paragraph{MBA obfuscation and deobfuscation.}
Mixed-Boolean Arithmetic was introduced as an obfuscation technique by~\citet{zhou2007information}, who observed that mixing bitwise and arithmetic operations creates expressions that are difficult to simplify.
\citet{eyrolles2017obfuscation} provided a comprehensive study of MBA obfuscation, including reconstruction and simplification tools.
MBA-Blast~\citep{liu2021mba} proposed a simplification method based on identifying the linear structure of MBA expressions, decomposing them into a combination of basis expressions.
GAMBA~\citep{reichenwallner2023gamba} extended this to general (not just linear) MBA expressions using an algebraic approach.
\citet{reichenwallner2022efficient} gave efficient algorithms specifically for linear MBA deobfuscation.
More recent work explores non-linear MBA obfuscation~\citep{liu2021nonlinear} and architectural hardening against deobfuscation~\citep{wei2026unifying}.
Formal verification of MBA obfuscation has been studied by~\citet{blazy2019formal}.

\paragraph{Program synthesis for deobfuscation.}
Syntia~\citep{blazytko2017syntia} pioneered the use of stochastic program synthesis (Monte Carlo Tree Search) for black-box deobfuscation of binary code, including MBA-obfuscated expressions.
Xyntia~\citep{MenguyBBL21a} studied, improved, and proposed mitigations for local search-based deobfuscation, establishing a stronger baseline in this line of work.
QSynth~\citep{david2020qsynth} refined this approach with better sampling and synthesis strategies.
\citet{lee2023simplifying} combined program synthesis with term rewriting for MBA deobfuscation.
XSmir~\citep{attias2025augmenting} augmented search-based synthesis with local inference rules to improve effectiveness.
NeuReduce~\citep{feng2020neureduce} took a neural approach, training a recurrent neural network to simplify MBA expressions, trading soundness for speed.

\paragraph{GPU-accelerated program synthesis.}
\citet{valizadeh2023paresy} demonstrated the first GPU-accelerated program synthesizer for regular expression inference, using bitvector representations and a language cache to achieve orders-of-magnitude speedups.
\citet{valizadeh2024ltl} extended this to LTL formula learning, introducing divide-and-conquer and relaxed uniqueness checks (RUCs) to scale further.
\citet{berger2025enumerate} presented a general framework for GPU-accelerated synthesis based on enumerating semantics rather than syntax.
All these works operate in qualitative domains where observational equivalence classes can be compactly represented and cached.
\SIMBA{} is the first to tackle a quantitative domain, requiring fundamentally new cache-free algorithms.

\paragraph{General program synthesis.}
Bottom-up enumerative synthesis~\citep{alur2013sygus, udupa2013transit} is a foundational technique that systematically generates programs of increasing size.
Observational equivalence reduction~\citep{udupa2013transit} is the standard pruning technique: only one representative per equivalence class is kept.
Syntax-guided synthesis (SyGuS)~\citep{alur2013sygus} provides a formal framework and annual competition.
State-of-the-art SyGuS solvers like cvc5~\citep{cvc5} use sophisticated enumeration strategies but are CPU-based and do not exploit GPU parallelism.

\section{Conclusion}
\label{sec:conclusion}
We presented \SIMBA{}, the first GPU-accelerated synthesizer for Mixed-Boolean Arithmetic expressions.
By introducing a cache-free enumeration strategy, we overcame the fundamental limitation of prior GPU-based synthesis approaches, which rely on language caches that are impractical for quantitative domains.
Our work opens the door to GPU-accelerated program synthesis for a wide class of quantitative domains beyond MBA, wherever the output space is too large for caching to be effective.

\paragraph{Limitations.}
\SIMBA{} has several limitations.
First, the approach requires a GPU: the algorithm is designed around massively parallel thread execution and provides no benefit on CPU-only hardware.
Second, despite the cache-free design, the number of candidate expressions still grows super-exponentially with formula size, so \SIMBA{} remains practical only up to sizes around 16 in our experiments; very large target expressions remain out of reach within reasonable time budgets.
Finally, \SIMBA{} does not handle approximate or noisy oracles, and cannot synthesize expressions with numerical constants.

\paragraph{Broader impacts.}
\SIMBA{} is intended for defensive security applications such as malware deobfuscation, compiler superoptimization, and reverse engineering.
Like any synthesis tool, it could in principle be used to generate or validate obfuscated code; however, obfuscation tools are already widely available, so we do not anticipate that \SIMBA{} meaningfully lowers that barrier.
We do not foresee significant negative societal impacts beyond those already present in the broader program analysis ecosystem.

\bibliography{references}
\bibliographystyle{plainnat}

\appendix
\section{Comparison with Existing Tools}
\label{app:comparison_existing_tools}

\begin{figure}[htbp]
  \begin{centering}
    \import{figures/}{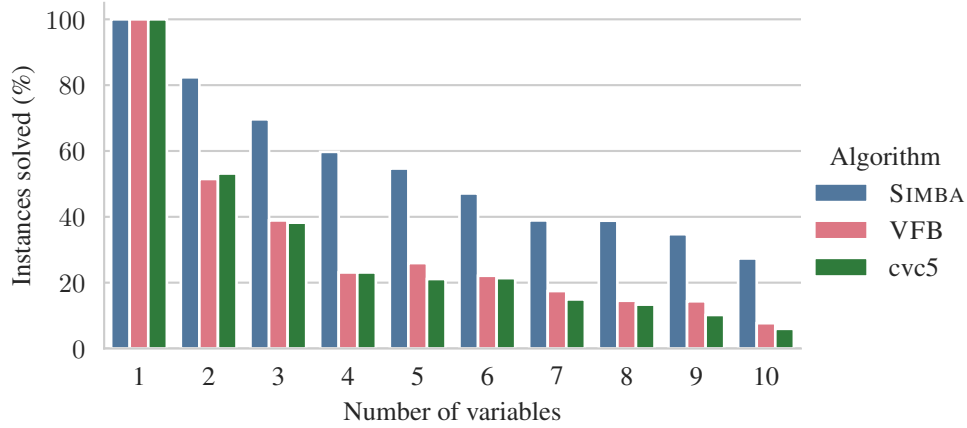}
    \caption{Percentage of instances solved by each algorithm across varying numbers of variables.}\label{fig:vars}
  \end{centering}
\end{figure}

\section{Cache Size and Efficiency Tables}
\label{app:cache-tables}

The following tables complement \cref{tab:cache-5} in the main body by reporting cache size and efficiency for \VFB across all variable counts $k \in \{3, 4, 6, 7, 8\}$.

\begin{table}[htbp]
\centering
\caption{Cache size and efficiency of \VFB for $k = 3$ variables.}
\label{tab:cache-3}
\begin{tabular}{rrrrrr}
\toprule
Size & \#MBA & \#VFB cache & VFB mem & \% cached & VFB cum.\ time (s) \\
\midrule
2 & 9 & 9 & $<$1\,MB & 100.0\% & 1.2 \\
3 & 75 & 38 & $<$1\,MB & 50.7\% & 2.4 \\
4 & 333 & 71 & $<$1\,MB & 21.3\% & 3.4 \\
5 & 2,451 & 554 & $<$1\,MB & 22.6\% & 4.3 \\
6 & 14,877 & 2,112 & $<$1\,MB & 14.2\% & 5.5 \\
7 & 121,179 & 21,981 & 1.4\,MB & 18.1\% & 6.5 \\
8 & 802,881 & 97,071 & 6.2\,MB & 12.1\% & 7.4 \\
9 & 6,298,419 & 756,167 & 48.4\,MB & 12.0\% & 8.5 \\
10 & 45,635,841 & 4,079,306 & 261.1\,MB & 8.9\% & 9.3 \\
11 & 366,816,891 & 30,968,139 & 2.0\,GB & 8.4\% & 10.0 \\
12 & 2,757,233,061 & OOM & OOM & OOM & OOM \\
\bottomrule
\end{tabular}
\end{table}

\begin{table}[htbp]
\centering
\caption{Cache size and efficiency of \VFB for $k = 4$ variables.}
\label{tab:cache-4}
\begin{tabular}{rrrrrr}
\toprule
Size & \#MBA & \#VFB cache & VFB mem & \% cached & VFB cum.\ time (s) \\
\midrule
2 & 12 & 12 & $<$1\,MB & 100.0\% & 1.2 \\
3 & 124 & 65 & $<$1\,MB & 52.4\% & 2.4 \\
4 & 572 & 114 & $<$1\,MB & 19.9\% & 3.4 \\
5 & 4,988 & 1,051 & $<$1\,MB & 21.1\% & 4.3 \\
6 & 32,636 & 4,086 & $<$1\,MB & 12.5\% & 5.5 \\
7 & 311,932 & 55,253 & 3.5\,MB & 17.7\% & 6.5 \\
8 & 2,243,196 & 244,245 & 15.6\,MB & 10.9\% & 7.4 \\
9 & 20,140,668 & 2,298,267 & 147.1\,MB & 11.4\% & 8.5 \\
10 & 161,176,188 & 13,036,731 & 834.4\,MB & 8.1\% & 9.3 \\
11 & 1,475,205,756 & 31,589,728 & 2.0\,GB & 2.1\% & 10.0 \\
12 & 12,299,156,092 & OOM & OOM & OOM & OOM \\
\bottomrule
\end{tabular}
\end{table}

\begin{table}[htbp]
\centering
\caption{Cache size and efficiency of \VFB for $k = 6$ variables.}
\label{tab:cache-6}
\begin{tabular}{rrrrrr}
\toprule
Size & \#MBA & \#VFB cache & VFB mem & \% cached & VFB cum.\ time (s) \\
\midrule
2 & 18 & 18 & $<$1\,MB & 100.0\% & 1.2 \\
3 & 258 & 140 & $<$1\,MB & 54.3\% & 2.4 \\
4 & 1,242 & 227 & $<$1\,MB & 18.3\% & 3.4 \\
5 & 14,154 & 2,676 & $<$1\,MB & 18.9\% & 4.3 \\
6 & 101,466 & 10,586 & $<$1\,MB & 10.4\% & 5.5 \\
7 & 1,246,650 & 212,216 & 13.6\,MB & 17.0\% & 6.5 \\
8 & 9,941,850 & 919,469 & 58.8\,MB & 9.2\% & 7.4 \\
9 & 110,265,882 & 11,546,211 & 739.0\,MB & 10.5\% & 8.5 \\
10 & 1,007,929,818 & 32,230,181 & 2.1\,GB & 3.2\% & 9.3 \\
11 & 11,272,896,474 & OOM & OOM & OOM & OOM \\
\bottomrule
\end{tabular}
\end{table}

\begin{table}[htbp]
\centering
\caption{Cache size and efficiency of \VFB for $k = 7$ variables.}
\label{tab:cache-7}
\begin{tabular}{rrrrrr}
\toprule
Size & \#MBA & \#VFB cache & VFB mem & \% cached & VFB cum.\ time (s) \\
\midrule
2 & 21 & 21 & $<$1\,MB & 100.0\% & 1.2 \\
3 & 343 & 188 & $<$1\,MB & 54.8\% & 2.4 \\
4 & 1,673 & 297 & $<$1\,MB & 17.8\% & 3.4 \\
5 & 21,287 & 3,860 & $<$1\,MB & 18.1\% & 4.3 \\
6 & 157,241 & 15,332 & $<$1\,MB & 9.8\% & 5.5 \\
7 & 2,142,679 & 359,149 & 23.0\,MB & 16.8\% & 6.5 \\
8 & 17,695,293 & 1,536,260 & 98.3\,MB & 8.7\% & 7.4 \\
9 & 214,233,831 & 21,655,933 & 1.4\,GB & 10.1\% & 8.5 \\
10 & 2,053,008,573 & OOM & OOM & OOM & OOM \\
\bottomrule
\end{tabular}
\end{table}

\begin{table}[htbp]
\centering
\caption{Cache size and efficiency of \VFB for $k = 8$ variables.}
\label{tab:cache-8}
\begin{tabular}{rrrrrr}
\toprule
Size & \#MBA & \#VFB cache & VFB mem & \% cached & VFB cum.\ time (s) \\
\midrule
2 & 24 & 24 & $<$1\,MB & 100.0\% & 1.2 \\
3 & 440 & 243 & $<$1\,MB & 55.2\% & 2.4 \\
4 & 2,168 & 376 & $<$1\,MB & 17.3\% & 3.4 \\
5 & 30,456 & 5,329 & $<$1\,MB & 17.5\% & 4.3 \\
6 & 230,392 & 21,222 & 1.4\,MB & 9.2\% & 5.5 \\
7 & 3,446,264 & 570,096 & 36.5\,MB & 16.5\% & 6.5 \\
8 & 29,274,616 & 2,407,325 & 154.1\,MB & 8.2\% & 7.4 \\
9 & 383,703,544 & 33,244,298 & 2.1\,GB & 8.7\% & 8.5 \\
10 & 3,823,512,056 & OOM & OOM & OOM & OOM \\
\bottomrule
\end{tabular}
\end{table}

\section{Hardware Comparison}
\label{app:hardware}

\begin{figure}[htbp]
  \begin{centering}
    \import{figures/}{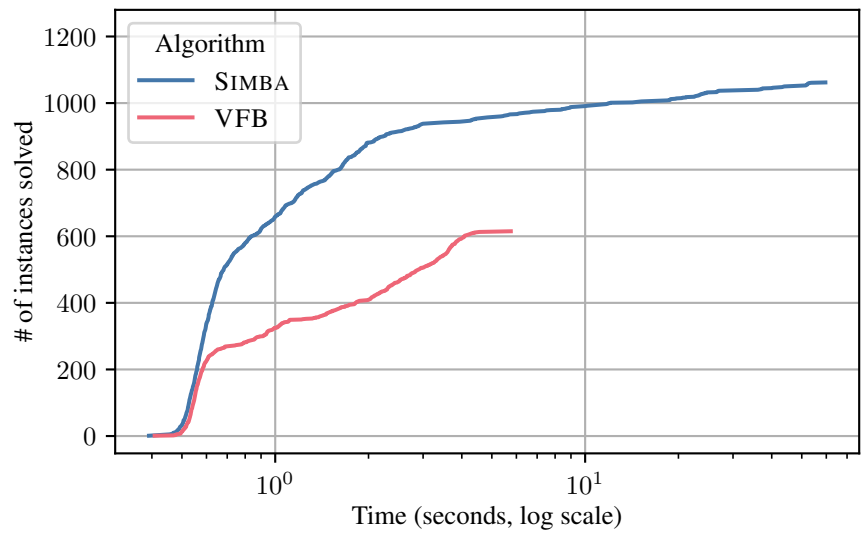}
    \caption{Number of instances solved under each time threshold (A100).}\label{fig:res-a100}
  \end{centering}
\end{figure}

\Cref{fig:res-h100,fig:res-a100} compare the number of instances solved within each time threshold by \SIMBA and \VFB on H100 and A100 GPUs respectively.
The results are consistent across both hardware configurations: \SIMBA substantially outperforms \VFB on both GPUs, solving roughly twice as many instances within the 60-second limit.
The absolute solving times are slightly faster on H100 GPUs than on A100 GPUs, as expected from the higher compute throughput of the H100, but the relative advantage of \SIMBA over \VFB is preserved across both platforms.

\begin{figure}[htbp]
  \begin{centering}
    \import{figures/}{res_h100.pgf}
    \caption{Number of instances solved under each time threshold (H100).}\label{fig:res-h100}
  \end{centering}
\end{figure}

\begin{figure}[htbp]
  \begin{centering}
    \import{figures/}{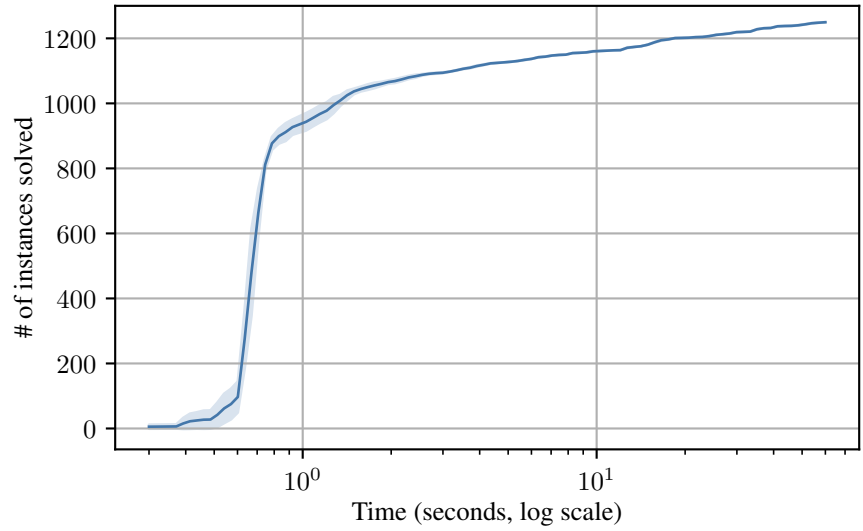}
    \caption{Number of instances solved by \SIMBA{} under each time threshold on H100 GPU, averaged over 5 independent runs. The shaded area shows $\pm$ standard deviation.}\label{fig:stability}
  \end{centering}
\end{figure}

\paragraph{Stability of \SIMBA.}
\Cref{fig:stability} reports the number of instances solved by \SIMBA within each time threshold, averaged over 5 independent runs on the H100 GPU.
The shaded band, representing $\pm$ one standard deviation, is barely visible throughout the curve, indicating that \SIMBA's performance is highly reproducible.
This is expected given that \SIMBA is a deterministic algorithm: the only source of variance is the scheduling jitter of the GPU runtime, which is negligible at the scale of our experiments.

\end{document}